\documentclass[aps,prd,a4paper,twocolumn,showpacs,showkeys,preprintnumbers,amsmath,amssymb,floatfix]{revtex4}
\usepackage{graphicx}
\usepackage{bm}
\usepackage{dcolumn}
\usepackage{color}
\usepackage{pst-plot}

\newcommand{\hide}[1]{{}}

    \def\be{\begin{equation}}
    \def\ee{\end{equation}}
    \def\ba{\begin{eqnarray}}
    \def\ea{\end{eqnarray}}

\newcommand{\lsim}{\,\raisebox{-0.6ex}{$\buildrel < \over \sim$}\,}
\newcommand{\gsim}{\,\raisebox{-0.6ex}{$\buildrel > \over \sim$}\,}

\begin{document}

\title{On Adiabatic Renormalization of Inflationary Perturbations}

\author{Ruth Durrer${}^{1}$\footnote{ruth.durrer@unige.ch}, Giovanni 
Marozzi${}^{2}$\footnote{marozzi@bo.infn.it}, and 
Massimiliano Rinaldi${}^{1}$\footnote{massimiliano.rinaldi@unige.ch}}

\affiliation{${}^{1}$Universit\'e de Gen\`eve, D\'epartment de Physique 
Th\'eorique, \\
24 quai Ernest Ansermet  CH--1211 Gen\`eve 4, Switzerland.\\
${}^{2}$ Dipartimento di Fisica dell'Universit\'a di Bologna and I.N.F.N.,  \\
Via Irnerio, 46, 40126 Bologna, Italy}

\date{\today}

\begin{abstract}
\noindent We discuss the impact of adiabatic renormalization on
the power spectrum of scalar and tensor perturbations from inflation. 
We show that adiabatic regularization is ambiguous as it 
leads to very different results, for different
adiabatic subtraction schemes, both in the range  $v\equiv k/(aH) \gsim 0.1$
and in the infrared regime. All these schemes agree in the far ultraviolet, $v\gg 1$.
Therefore, we argue that  in the far infrared regime, 
$v\ll 1$,  the adiabatic expansion is no longer valid,
and the unrenormalized spectra are the physical, measurable quantities.  
These findings cast some doubt on the validity of the adiabatic 
subtraction at horizon exit, $v=1$, to determine the perturbation spectra 
from inflation which has recently advocated in the literature.
\end{abstract}


\maketitle

\section{Introduction}
\label{one}

\noindent Inflation was originally proposed to solve the initial condition 
problem of 
standard big bang cosmology. At the same time, it was found
that inflation typically leads to a nearly scale invariant spectrum of scalar 
and tensor fluctuations~\cite{Muk,Star}. It is this  finding, which is so well 
confirmed
by the observed anisotropies and polarization in the cosmic microwave 
background~\cite{WMAP5}, which has led to a wide acceptance of the 
inflationary paradigm. Using present and future CMB data, in combination with
other cosmological data sets, we are now in the position to constrain 
models. 

Typically, an inflationary model predicts the value of three parameters, 
namely the scalar spectral index $n_s$,
the tensor spectral index $n_t$, and
the tensor-to-scalar ratio $r$. So far,  observations just provide upper limits on tensor 
fluctuations. These are not independent of the scalar spectral index $n_s$ 
as it is evident from the 2-dimensional one- and two-$\sigma$ confidence 
contours, shown in Fig.~5 of Ref.~\cite{WMAP5para}. These data can be used to
constrain inflationary models. For example, in~\cite{WMAP5para} it is noted
that a model of inflation with a scalar field potential of the form 
$\lambda\phi^4$ is ruled out if the number $N$ of e-foldings of 
inflation after horizon crossing of the scales probed by WMAP, 
$k\simeq 0.002h/$Mpc $\simeq 6H_0$, is of the order of $N\lsim 50$ -- $60$. 
In this expression,
$H_0$ is the current value of the Hubble parameter and $h=0.72\pm0.08$.

This is a truly breath taking result meaning that CMB data, i.e. cosmological 
observations on the largest scales, can provide information about the physics 
at energy 
scales much higher than those attainable in the laboratory, hence about
the physics on the smallest scales. It is therefore of the utmost importance 
that these 
results are subjected to the deepest scrutiny. With this point in mind,  we
have studied the recent works \cite{Parker1}-\cite{ParkeretAll}. In particular,
in Ref.~\cite{Parker1}, the author argues that the inflationary power
spectra, as they are usually calculated, are not correct. In fact, since they 
diverge 
at coincident points, one should subtract an appropriate adiabatic 
counter\-term
(see also \cite{FMVV3} for a different point of view).
In Ref. \cite{ParkeretAll} the authors perform explicit calculations along 
these lines, and  
 subtract the adiabatic term at the Hubble exit, namely when 
$v=k/(aH)=1$. As a result, the values of the tensor-to-scalar ratio that they find, differ 
significantly from the ones usually adopted to be compared with the data, 
\cite{WMAP5para}.
The most surprising consequence is that, for example, the chaotic inflationary 
model $\lambda\phi^4$ is no longer ruled out 
by the WMAP data.

It is well known that the standard power spectra are nearly time-independent 
on super-Hubble scales, i.e. when $v\leq1$. 
In this paper we show that this is not the case for the adiabatic contribution
to the spectra in a realistic model of inflation. 
The renormalized spectra, no matter the approach used for the adiabatic subtraction,
always depend on $v$. The adiabatic 
regularization, even when performed at horizon exit or a few Hubble times later, 
is ambiguous in the sense that it gives different result depending on  which approach is used. In some cases, the  
 result is even strongly time-dependent. In fact, we show that there are 
different ways to perform the adiabatic 
subtraction, which all agree in the far ultraviolet regime, but yield very 
different results for $v\leq 1$.
The most reasonable adiabatic expression actually yields an adiabatic 
spectrum $P^{(2)}$ such that the ratio   $P^{(2)}/P^{(IR)}$ becomes quickly negligible 
for $v\ll 1$, indicating that the spectrum is not modified.

These considerations indicate that the correct time at which the 
adiabatic subtraction has to be performed is the end of inflation, rather than
the time of Hubble exit. However, at the end of inflation all the modes 
relevant for observational cosmology are in the far infrared region where
the adiabatic expansion seems not appropriate as the expansion of the 
Universe is not slow compared to the oscillation frequency of the mode and, 
thus, not adiabatic. If one insists, however, 
to extend the adiabatic regularization in the infrared, we present an argument, 
which shows that the adiabatic counterterms become negligible anyway.
Thus, we shall argue that the
physical result is not affected by adiabatic regularization.
 
The paper is organized as follows: in the next section we present approximate 
expressions for the scalar and tensor power spectra in the framework 
of slow-roll inflation. In Section~\ref{four} we discuss different adiabatic 
subtractions for both the scalar and tensor power spectrum, and we argue 
that the difference with the original power spectra becomes irrelevant in 
the far
infrared regime. In Section~\ref{five} we draw our conclusions. 
Some technical results are deferred to appendices.

\section{Power spectra from slow-roll inflation}
\label{two}

\subsection{Linear perturbations in slow-roll inflation}

\noindent We consider a spatially flat Universe, whose dynamics is driven 
by a classical minimally coupled scalar field, described by the action 
    \be
    S = \int d^4x \sqrt{-g} \left[ 
\frac{R}{16{\pi}G}
    - \frac{1}{2} g^{\mu \nu}
    \partial_{\mu} \phi \partial_{\nu} \phi
    - V(\phi) \right]\ .
    \label{action}
\ee
For a spatially flat Friedmann spacetime, of the form 
$ds^2=-dt^2+a^2(t)\delta_{ij}dx^idx^j$, the background equations of motion 
for $\phi$ and for the scale factor $a(t)$ read
\begin{eqnarray}
&& \ddot{\phi}+3H\dot{\phi}+V_\phi=0 \label{eqmotion1}\ ,  \\
&& \left(\frac{\dot a}{a}\right)^2=H^2 = \frac{1}{3 M _{\rm pl}^2} 
        \left[ \frac{\dot \phi^2}{2} + V \right]\ , \label{Fried}\\
&& \dot{H}=-\frac{1}{2 M _{\rm pl}^2} \dot{\phi}^2\ ,
\label{EE00}
\end{eqnarray}
where $M_{\rm pl}^2=1/(8\pi G)$ is the reduced Planck mass. 
The dot denotes a derivative with respect to the cosmic time $t$.
 Linear perturbations of the metric in longitudinal gauge are given by
\be
ds^2 = -(1+2\Psi)dt^2 +a^2\left[(1-2\Phi)\delta_{ij} + h_{ij}\right]dx^idx^j\ ,
\ee 
Here $\Psi$ and $\Phi$ represent the Bardeen potentials, and $h_{ij}$ describes 
traceless, transverse tensor degrees of freedom, that is gravitational waves.
We do not discuss vector perturbations.

In single-field inflationary models, and to first order in perturbation 
theory, we have $\Phi=\Psi$.  Scalar perturbations have 
only one 
degree of freedom, which can be studied by means of a single gauge 
invariant variable, such as 
the so-called Mukhanov variable \cite{MukhanovV}, defined, 
in longitudinal gauge, by
\be
Q=\varphi+\frac{\dot{\phi}}{H}\Psi \ .
\label{MVariable}
\ee
In this expression, we assume that the scalar field can be written as a 
background term plus a linear perturbation, namely as $\phi+\varphi$.

Often, one also uses the curvature variable $\zeta$, which, for $\Phi=\Psi$,
is defined as~\cite{mybook}
\be
\zeta = \frac{H}{\dot\phi}Q = \frac{2(H^{-1}\dot\Psi + \Psi)}{3(1+w)} +\Psi\ .
\ee
Here, $w$ is the equation of state parameter, which satisfies
\be
1+w=
\frac{2\dot\phi^2}{\dot\phi^2 +2V(\phi)}\ .
\ee

It is important to note that both $\zeta$ and $Q$ are related to the Bardeen 
potential $\Psi$ via a first order equation. They are not independent degrees 
of freedom and
it is therefore not consistent to think of $Q$ as a quantum degree of freedom 
and of $\Psi$ as a classical variable. When we quantize $Q$, or rather $aQ$
as below, we also quantize the Bardeen potential. In fact, we 
do not equate expectation values of some quantum fields to 
classical first order perturbations of the metric via Einstein's equation, 
but we do quantize the metric perturbations. 

The equation governing $Q$ in Fourier space, is given by
\be
\ddot{Q}_k + 3 H \dot{Q}_k + \frac{1}{a^2}k^2 Q_k +
\left[ V_{\phi \phi} + 2 \frac{d}{dt}\left(3 H + 
\frac{\dot H}{H}\right)\right] Q_k = 0 \ ,
\label{Eq_mukhanov}
\ee
where $V_{\phi\phi}$ denotes the second derivative of $V$ with respect to $\phi$.
During inflation, we assume that the so-called slow-roll parameters
\be
\epsilon=\frac{M_{\rm pl}^2}{2} \left(\frac{V_\phi}{V}\right)^2\ , \qquad
\eta=M_{\rm pl}^2 \frac{V_{\phi\phi}}{V}\ 
\label{SRParameter}
\ee
are small, $\epsilon, ~ |\eta| \ll 1$. To leading order in these 
parameters, each mode $Q_k$ satisfies the equation
\be
\ddot{Q}_k + 3 H \dot{Q}_k +H^2 \left[ \frac{k^2}{a^2 H^2} + 3 \eta-6 \epsilon
\right] Q_k = 0 \ .
\label{Eq_mukhanov_Fourier}
\ee

Analytic solutions for Eq.~(\ref{Eq_mukhanov_Fourier}) can be found  
if the slow-roll parameters are constant. More generally, 
along the lines of \cite{FMVV,FMSVV}, we can study this equation in the 
infrared regime (IR), 
corresponding to $k/(aH)< c\ $, and in the ultraviolet regime (UV), 
corresponding to $k/(aH)>c$ , where $1/10\lesssim c \lesssim 1$.
We will shortly see that we can ``match'' the UV solution to the IR one at 
$k/(aH)=c$.
In the UV, the slow-roll parameters can be considered as constant. The 
canonically normalized solution to Eq.\ (\ref{Eq_mukhanov_Fourier}) 
with adiabatic vacuum initial conditions then reads
\be 
Q^{(UV)}_k=\frac{1}{a^{3/2}} \sqrt{\frac{\pi (1+\epsilon)}{4 H}}
H_\nu^{(1)}\left[\frac{k}{a H} (1+\epsilon)\right]\ ,
\label{solution_Mukhanov_UV}
\ee
where  $H_\nu^{(1)}$ is the Hankel function of the first kind with index 
$\nu=\frac{3}{2}-\eta+3 \epsilon$.
It is instructive to rewrite Eq.\ (\ref{Eq_mukhanov}) in the form
\ba
(a Q_k)'' + \left( k^2 - \frac{z''}{z} \right) a Q_k &=& 0 \,,  
\label{Eq_canonical_mukhanov}\\
\mbox{where} \quad z = a \frac{\dot \phi}{H} = -aM_{pl}\sqrt{2\epsilon}\ , &&
\ea
and primes denote derivatives with respect to conformal time $\tau$, defined 
by $ad\tau =dt$. Eq.~(\ref{Eq_canonical_mukhanov}) is simply 
the equation of
a harmonic oscillator with a negative time-dependent mass $-z''/z$.
When $ k^2 - \frac{z''}{z}<0$, this leads to amplification on the mode $aQ_k$.
This form of the perturbation equation is completely general and independent 
of the form of the potential.

In the far IR, $v\ll 1$, one can neglect the term $k^2$, and 
the non-decaying mode of the solution $Q_k$ is well approximated by 
$Q_k \propto \frac{\dot{\phi}}{H} =-M_{pl}\sqrt{2\epsilon}$.
On the other hand, in the far UV, $-k\tau \simeq v \gg 1$ and one can neglect 
the term $z''/z$, so that Eq.~(\ref{Eq_canonical_mukhanov}) reduces to the 
equation for a simple harmonic oscillator. 

As mentioned above, by imposing that the UV solution approximately matches 
the IR solution for 
$k/(a H)=c$, we obtain the solution valid for $k<caH$, namely \cite{FMSVV}
\be 
Q^{(IR)}_k=\frac{1}{a^{3/2}} \sqrt{\frac{\pi(1+\epsilon)}{4 H} }
\left(\frac{H_c}{H}\right)^{\gamma} 
H_{3/2}^{(1)}\left[\frac{k}{a H}(1+\epsilon)\right]\ .
\ee
Here, $H_c$ is the value of the Hubble parameter at the time $t_c$ when 
$k=caH$, and 
\be
\gamma=3+{V_{\phi\phi}\over 3\dot{H}}  = 3\left(1 
-\frac{\eta}{3 \epsilon}\right) \ .
\ee

We now turn to tensor perturbations $h_{i j}$. In Fourier space both
tensor polarizations evolve according to
\be
\ddot h_{ k} + 3 H \dot{h}_{ k} + \frac{k^2}{a^2} 
h_{ k} = 0\ .
\label{Eq_graviton_Fourier}
\ee
As before,  one can derive approximate solutions in the UV and IR. The 
mode and the amplitude are chosen such that the canonically normalized 
variable $(aM_{\rm pl}/\sqrt{2})h$ satisfies adiabatic vacuum initial 
conditions in the UV. Thus, one finds
\be 
h^{(UV)}_k=\frac{1}{a^{3/2}M_{\rm pl}^2} \sqrt{\frac{\pi(1+\epsilon)}{2 H}}
H_\nu^{(1)}\left[\frac{k}{a H}(1+\epsilon)\right]\ ,
\ee
with $\nu=3/2+\epsilon$, and
\be 
h^{(IR)}_k=\frac{1}{a^{3/2}M_{\rm pl}^2} \sqrt{\frac{\pi(1+\epsilon)}{2 H}}
\left(H_c\over H\right)
H_{3/2}^{(1)}\left[\frac{k}{a H}(1+\epsilon)\right] .
\label{solution_tensor_UV}
\ee

\subsection{Power Spectra}
\label{three}
\noindent With the solutions discussed in the previous subsection, we can now 
compute the scalar and tensor power spectrum, defined by
\be
P_{\zeta}(k)=\frac{k^3}{2 \pi^2} \left(\frac{H}{\dot{\phi}}\right)^2 |Q_k|^2\ ,
\quad P_{t}(k)=\frac{2 k^3}{\pi^2}  |h_k|^2\ .
\label{spectrum}
\ee
In terms of the variable $v=k/(a H)$, the expansion of the spectra to first 
order in the slow-roll parameters, in the UV and IR, respectively yields
\ba
P^{(UV)}_{\zeta}&=&\frac{1}{2 M_{\rm pl}^2} \left(\frac{H_v}{2 \pi}\right)^2 
\left[1+v^2+f(v) \epsilon_v+g(v) \eta_v\over\epsilon_v \right],\nonumber 
\\ \\
P^{(UV)}_{t}&=&\frac{8}{M_{\rm pl}^2}\left(\frac{H_v}{2 \pi}\right)^2 
\left[1+v^2+f_t(v) \epsilon_v\right]\ ,
\ea
and 
\ba\label{PIRR}
P^{(IR)}_{\zeta}&=&\frac{1}{2 M_{\rm pl}^2} \left(\frac{H_v}{2 \pi}
\right)^2  \left(\frac{H_c}{H_v}\right)^{2 \gamma} 
\left[1+v^2-2 \epsilon_v\over \epsilon_v\right]\ , \nonumber\\ \\
P^{(IR)}_{t}&=&\frac{8}{M_{\rm pl}^2}\left(\frac{H_v}{2 \pi}\right)^2 
\left(\frac{H_c}{H_v}\right)^2
\left[1+v^2-2 \epsilon_v\right]\ .\label{PIRt}
\ea
These UV and IR spectra have the correct asymptotic form but they do not 
match 
exactly at $v=c$ since we have neglected the decaying mode contribution in 
$P^{(IR)}$. The small discontinuity is of the order of the slow-roll 
parameters.
The functions $f(v)$, $f_t(v)$, and $g(v)$ appearing in these expressions are 
defined in Appendix \ref{defin} and plotted in Fig. \ref{figFunction}.
$H_v$, $\epsilon_v$, and $\eta_v$ are the values of these quantities 
calculated at the time $t_v$ for which $k/(a H)=v$. 
The functions  $f(v)$, $f_t(v)$, and $g(v)$  always appear multiplied by 
$\eta$ or $\epsilon$. 
Therefore,  they are always subdominant for wave numbers $k$ which exit the 
Hubble scale in the 
slow-roll regime, i.e. when $\epsilon_v\ll 1$ and $\eta_v\ll 1$.

An important  observable parameter is the tensor-to-scalar ratio 
$r=P_t/P_\zeta$. 
On considering the particular case when
$V=m^2 \phi^2/2$, we have $\eta=\epsilon$ and $\gamma=2$.
Therefore, for this particular case, and at the leading order in the slow-roll 
parameters, we find
\be
r^{(UV)}=16 \epsilon_v\ , \quad r^{(IR)}=16 \epsilon_c\ .
\label{ratios}
\ee  
Note that, during slow-roll evolution, $\epsilon$ varies slowly 
($\dot\epsilon$  is second order in the slow-roll parameters), so that  
also $r^{(IR)}$ is nearly constant for scales which reach $k=c a H$ during
slow-roll.

To compare our findings with the five-year WMAP results, we must 
write $r^{(UV, IR)}$ in terms of the spectral index 
$n_s=1+\frac{d}{d\ln k}\ln P_\zeta$. In turn, $n_s$ must be expressed as a 
function of $v$ and $N$,  i.e. the number of $e$-folds between the 
epoch when the modes corresponding to the scales probed by WMAP 
exit the Hubble scale and  the end of inflation. 
In Appendix \ref{spectra}, we show that, when  
$V=m^2 \phi^2/2$,
\ba\label{e:ns}
n_s^{(UV)}=1-4 \epsilon_v\ , \quad n_s^{(IR)}=1-4 \epsilon_c\ ,
\ea
where 
\be
\epsilon_s=\eta_s=\frac{1}{2 (N+ \ln s)}\ .
\ee
while $s$ is either $v$ or $c$. It then follows that 
$$ r^{(UV,IR)}=4\left(1-n_s^{(UV,IR)}\right) \ .  $$

The generic slow-roll expression for the scalar spectral index is~\cite{mybook}
\be
n_s =1 -6\epsilon+2\eta\ .
\ee 
One easily verifies that for general chaotic inflation models with 
$V=\frac{\lambda}{p} \frac{\phi^p}{M_{pl}^{p-4}}$ 
one has $
\epsilon =\frac{p^2}{2}\left(\frac{M_{\rm pl}}{\phi}\right)^2$, 
while $\eta =p(p-1)\left(\frac{M_{\rm pl}}{\phi}\right)^2$, hence
\be
 n_s =1 -\left(2+\frac{4}{p}\right)\epsilon\ .
 \ee


\section{Renormalized power spectra}
\label{four}
\noindent We now investigate how the power spectrum is modified, when 
corrected by the subtraction of the adiabatic expansion up to the
second order. 

Let us briefly review how to obtain the adiabatic contribution 
to the power spectrum, in terms of  the Mukhanov variable, for 
the scalar perturbations.
As explained in the appendix of  \cite{Marozzi}, it is more convenient to 
formulate the adiabatic expansion by using the modulus of the Mukhanov variable
$x_k =\sqrt{2} |Q_{\bf k}|$, which satisfies
the Pinney equation
\be
\ddot{x}_k+3 H \dot{x}_k+
\left [\frac{k^2}{a^2}+
 V_{\phi \phi} + 2 {d\over dt}\left(3 H + 
\frac{\dot H}{H}\right)
\right ] x_k=\frac{1}{a^6 x^3_k} \,.
\label{EQMPF_k_x}
\ee
In conformal time, the above equation simplifies to
\ba
(a x_k)''+\Omega^2(a x_k)={1\over (a x_k)^3}\ ,
\label{conformal_eq_x_k}
\ea
where
\ba
\Omega_k^2=k^2+a^2 V_{\phi\phi}-{1\over 6}a^2\tilde{R}\ ,
\label{conformal_freq_x_k}
\ea
and
\be
\tilde{R} = R - 6 \left(-4 \frac{a^{'2}}{a^4}-2 \frac{a^{''2}}{a^2 a^{'2}}
+ 2 \frac{a^{'''}}{a^2 a^{'}}\right) ,\quad R=6{a''\over a}\ .
\label{ricci_general}
\ee
From these equations, one obtains the WKB expansion for $x_k$ up to the second 
adiabatic order, which reads
\be
x_k=\frac{1}{a}\frac{1}{\Omega_k^{1/2}}
\left( 1+\frac{1}{8}\frac{\Omega_k^{''}}{\Omega_k^3}-\frac{3}{16}
\frac{\Omega_k^{'2}}{\Omega_k^4} \right)\ ,
\label{fourth_1}
\ee
In turn, from this expression one immediately finds the second order adiabatic 
expansion of $|Q_k|^2$, namely 
\be
|Q_{k}|^2=\frac{1}{2 a^2\Omega_k}\left(1
+\frac{1}{4}\frac{\Omega_k^{''}}{\Omega_k^3}-\frac{3}{8}
\frac{\Omega_k^{'2}}{\Omega_k^4} \right)\ .
\label{fourth_2}
\ee
The term $V_{\phi\phi}$ could be considered of adiabatic order zero in $\Omega_k^2$, but, by 
using the field equations together with Eq.\ (\ref{SRParameter}), 
one obtains a different conclusion, namely that $V_{\phi\phi}=3 H^2 \eta$, which is in general of adiabatic order 
two as well as $\tilde{R}$ (see \cite{Marozzi} for a different 
interpretation). Below, we briefly discuss the case when $V_{\phi\phi}$ is 
considered as of order zero. In terms of the slow-roll parameters 
we have, at leading order, 
\ba
R&=&6H^2(2-\epsilon)\ , \quad \tilde R=6H^2(2+5\epsilon)\ , \\\nonumber\\
\Omega^2_k&=&k^2+a^2H^2(3\eta-5\epsilon-2)\ .
\ea 
Since the time dependence of $\Omega_k^2$ is already of second order, any derivative 
of $\Omega_k$ generate terms of adiabatic order greater than two. Thus, we 
can neglect the derivatives in Eq.~(\ref{fourth_2}), and the power spectrum 
to second adiabatic order is simply
\ba
P^{(2)}_{\zeta}&\equiv& {k^3\over 2\pi^2}\left(H_v\over \dot 
\phi\right){1\over 2a^2\Omega_k}\\\label{PZfull}
&=&{1\over 2M_{\rm pl}^2}\left(H_v\over 2\pi\right)^2{1\over \epsilon_v}\,
{v^3\over\sqrt{|2-v^2-3\eta_v+5\epsilon_v|}}\ ,\nonumber 
\\\label{Pscalar}
\ea
where we used the relation $(\dot \phi/H_v)^2=2M_{\rm pl}\epsilon_v$. The 
absolute value is necessary because, for $v^2<2-3\eta_v+5\epsilon_v\simeq 2$, 
one obtains a negative $\Omega_k^2$. The divergence at 
$v^2=2-3\eta_v+5\epsilon_v$ is not relevant in the far IR, $v\ll 1$ or the 
far UV, $v\gg 1$, but it indicates that one cannot trivially connect these two 
regions.

The standard adiabatic expansion then goes on to expand Eq.~(\ref{PZfull}) 
again up to second order (see, for example, 
\cite{Marozzi}), thus the spectrum reads
\ba\label{PZUV}
P^{(2)}_{\zeta}\simeq{1\over 2M_{\rm pl}}\left(H_v\over 2\pi\right)^2{1\over 
\epsilon_v}\left( 1+v^2-{3\over 2}\eta_v+{5\over 2}\epsilon_v\right)\ .
\ea
This expansion is clearly meaningful only in the far UV, i.e. for $v\gg1$.
If one wants to extend the validity of adiabatic renormalization to the 
IR, one should use Eq.\ (\ref{Pscalar}), which, however,
becomes  rapidly negligible, with respect to  $P_\zeta^{(IR)}$, when $v\ll 1$.

The adiabatic expansion for the tensor perturbation can be found using the 
results in Appendix A of~\cite{FMVV2} with $m^2=0$.
Here, the authors consider a scalar field propagating on an 
unperturbed space-time. On such a  background, the equation of motion of
the scalar field with $m^2=0$ coincides exactly with the equation of motion 
of the tensor perturbation, and one obtains directly
\ba
x_k={1\over a\sqrt{k^2-{1\over 6}a^2R}}\ ,
\ea
where now $x_k=|h_k|M_{\rm pl}$. 
Thus, the tensor adiabatic power spectrum reads
\be
P^{(2)}_{t} = \frac{8}{M_{\rm pl}^2} \left(\frac{H_v}{2 \pi}\right)^2 
 {v^3\over \sqrt{|2-v^2-\epsilon_v|}}\ .
\label{PTfull}
\ee
As above, in the far UV one can expand for large $v$, up to second abiabatic order, 
and find for the adiabatic spectrum
\be
P^{(2)}_{t} \simeq \frac{8}{M_{\rm pl}^2} \left(\frac{H_v}{2 \pi}\right)^2 
\left(v^2+1-\frac{\epsilon_v}{2}\right).
\label{PTUV}
\ee
With similar consideration as above, one sees that $P^{(2)}_{t}$, when 
expressed as in Eq.(\ref{PTfull}), becomes rapidly negligible respect $P_t^{(IR)}$ when 
$v\ll 1$. 

\begin{figure}[!h]
\includegraphics[width=0.46\textwidth]{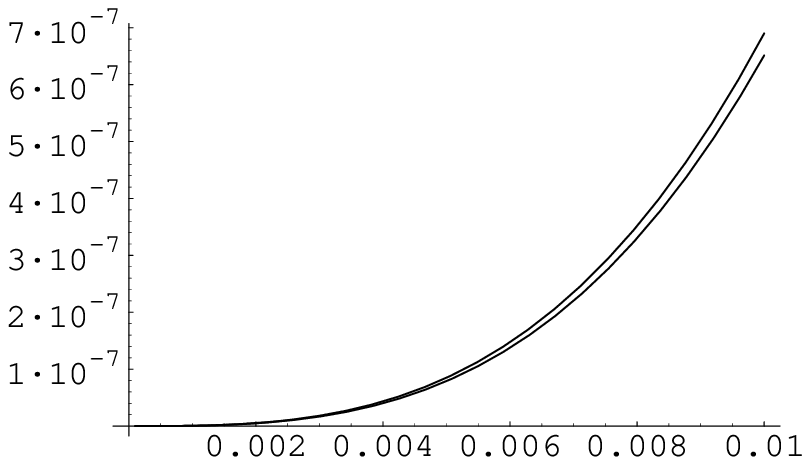}
\rput[tl](0,0.7){\Large $v$}
\rput[tl](-2.5,1.8){$P^{(2)}_\zeta/P^{(IR)}_\zeta$}
\rput[tl](-3.5,2.8){$P^{(2)}_t/P^{(IR)}_t$}
\caption{The ratios $P_{\zeta}^{(2)}/P_{\zeta}$ and $P_{t}^{(2)}/P_{t}$ in the 
IR, $v\ll 1$, calculated with Eqs. (\ref{PZfull}) and (\ref{PTfull}), for the
case $V=m^2\phi^2/2$. These contributions are negligible for both, scalar and 
tensor perturbations.
\vspace{0.3cm}}
\label{A}
\includegraphics[width=0.46\textwidth]{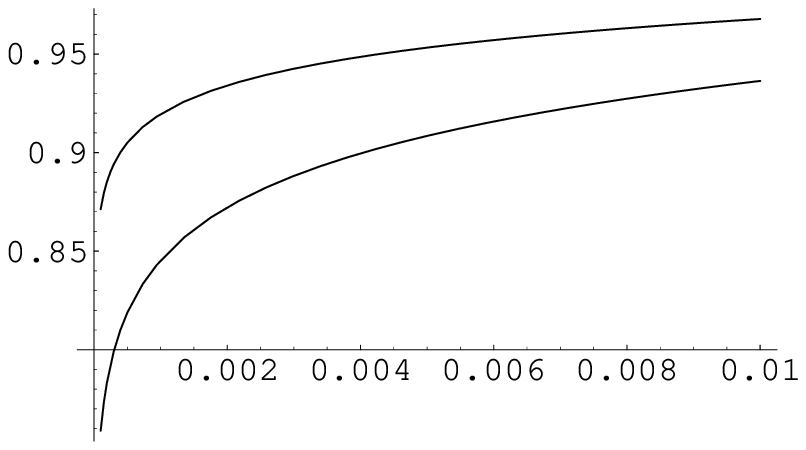}
\rput[tl](0,1.4){\Large $v$}
\rput[tl](-2.5,3.7){$P^{(2)}_\zeta/P^{(IR)}_\zeta$}
\rput[tl](-3.5,5.1){$P^{(2)}_t/P^{(IR)}_t$}
\caption{The ratios $P_{\zeta}^{(2)}/P_{\zeta}$ and $P_{t}^{(2)}/P_{t}$ in the 
IR, $v\ll 1$, calculated with Eqs. (\ref{PZUV}) and (\ref{PTUV}), for the 
case $V=m^2\phi^2/2$. These contributions are of considerable size for a large range of $v$, however they are not 
correct as in this regime the expansions (\ref{PZUV}, \ref{PTUV})  are not
 valid.}
\label{B}
\end{figure}

These results show that the extension to the IR of the adiabatic expansions 
of the scalar and tensor spectra, if sensible,  must be considered with great 
care. In fact, in the figures (\ref{A}) and (\ref{B}), one sees that the 
adiabatic corrections have very different effects, according to whether one uses Eqs.\ (\ref{PZfull}, \ref{PTfull}) or
Eqs.\ (\ref{PZUV}, \ref{PTUV}). 


We now consider the UV case in more detail. The renormalized power spectra are 
given  by
\be
P_\zeta=P_\zeta^{(UV)}-P_\zeta^{(2)}\ ,\quad P_t=P_t^{(UV)}-P^{(2)}_{t}\ ,
\ee
where $P_\zeta^{(2)}$ and $P^{(2)}_{t}$ can be calculated using 
either Eqs.\ (\ref{PZUV}, \ref{PTUV}) or Eqs.\ (\ref{PZfull}, \ref{PTfull}), 
as these fully coincide for large $v$. However, when $v=1$, the corresponding 
expressions for $P_\zeta$ already differ substantially. In fact, 
to leading order in the slow-roll parameters, using 
(\ref{PZUV}), we find
\be
P_\zeta\simeq\frac{1}{2 M_{\rm pl}^2} \left(\frac{H}{2 \pi}\right)^2 
 \frac{1}{\epsilon} \left[3 \alpha \epsilon+\left(3 \beta-\frac{9}{4}\right) 
\eta \right]\ ,
\ee
while, with Eq.(\ref{PZfull}), we obtain
\be
P_\zeta\simeq\frac{1}{2 M_{\rm pl}^2} \left(\frac{H}{2 \pi}\right)^2 
 \frac{1}{\epsilon} \left[1+\left(3 \alpha+5\right) \epsilon+\left(3 \beta-
\frac{21}{4}\right) 
\eta \right]\ ,
\ee
where $\alpha\simeq 0.903$, $\beta\simeq 0.449$, and all the parameters are 
calculated at the time when $k=a H$. These results make it clear that 
adiabatic counterterms, which agree in the far UV and correctly renormalize the 
theory in this regime, produce different results not only in the far IR, 
{\em but also at horizon exit}. Similar considerations hold for the tensor 
spectrum.

To better compare our results with the ones in \cite{ParkeretAll}, we now 
discuss the case where one considers $V_{\phi\phi}$ in the scalar spectrum 
of adiabatic order zero and  $\tilde{R}$ of order two. Starting from 
Eq.~(\ref{fourth_2}), and expanding again up to second order, one obtains 
\ba\nonumber
&&P^{(2)}_{\zeta}=\frac{1}{2 M_{\rm pl}^2} \left(\frac{H_v}{2 \pi}\right)^2 
\frac{1}{\epsilon_v} \Bigg[ \frac{v^3}{(v^2+3 \eta_v)^{1/2}}+\\
&&+\frac{v^3\left(1+\frac{5}{2}
\epsilon_v\right)}{(v^2+3 \eta_v)^{3/2}}+ \frac{9\eta_v v^3}{4(v^2+3 \eta_v)^{5/2}}-
\frac{45\eta_v^2 v^3}{8(v^2+3 \eta_v)^{7/2}} \Bigg],\nonumber\\
\label{PR_adiabatic}
\ea
where we have kept only the leading order, in the slow-roll parameters, in 
each term. With this expression, the renormalized $P_\zeta$ differs again 
from the other two expressions above. To see this, we set $v=1$ and expand 
with respect to the slow-roll parameters to find the expression
\be
P_\zeta=\frac{1}{2 M_{\rm pl}^2} \left(\frac{H}{2 \pi}\right)^2 {1\over \epsilon}
 \left(3 \alpha \epsilon+3 \beta \eta\right)\ ,
\ee
which has to be compared to Eq. (10) of \cite{ParkeretAll}. In this paper,
there is just $\alpha$ instead of $3\alpha$. This comes from the fact that we
use the Mukhanov variable, while in \cite{ParkeretAll} the authors use the 
scalar inflaton perturbation in a space-time without metric fluctuations.
However, when $v \ll 1$, this result is no longer reliable as we simply cannot 
expand with respect to the slow-roll parameter since
$3 \eta_v$ is no longer much smaller than $v^2$. For example, in the case
$V=m^2\phi^2/2$, as shown in Appendix \ref{spectra},
we have $3\epsilon_v\equiv3 \eta_v \simeq 1/(2 N)$. So, if $N=50$ and $v=1/5$, 
then $v^2=1/25$ and $3 \eta_v=3/100$.
As a result of this, the tensor-to-scalar ratio strongly depends on $v$ 
for $v<1$.

In addition to the above considerations, it is know that the long wave
adiabatic  modes do not contribute significantly to the renormalized
Green function at coincident points, in realistic inflationary models
(see \cite{KLPS} for similar consideration on massless particle).

The main point of this section is that adiabatic subtraction is not a 
reliable technique in the IR. In fact, we have presented three types of 
counterterms for the scalar spectrum, Eqs.\ (\ref{PZfull}), (\ref{PZUV}), 
and  (\ref{PR_adiabatic}), and two for the tensor one, namely 
Eqs.\ (\ref{PTfull}) and (\ref{PTUV}). In each case, these terms are 
equivalent in the far UV but give very different results elsewhere.
Furthermore, we have shown that when using (\ref{PZfull}) and 
(\ref{PTfull}), the subtracted spectra are close to the bare ones already 
at $v=1$ and virtually identical to them at $v\ll1$.

\section{Conclusions}

\label{five}

\noindent In this paper we have first determined the renormalized perturbation 
spectra in a slow-roll inflationary model in UV domain. In agreement 
with Ref.~\cite{ParkeretAll}, we have found that adiabatic subtraction
can lead to a substantial reduction of power in the UV. Namely, the 
larger $k/a$ in comparison to $H$, the closer the weight of the adiabatic 
counterterm becomes to the unrenormalized spectra. This is reasonable, 
since we do not expect that the expansion of the Universe is  
``energetic enough'' to excite physical modes in the UV. 

On the contrary,  in the IR, the adiabatic 
expansion is no longer valid, and 
there is no convincing physical argument to subtract  
this term to the standard power spectrum.
As there is a natural IR cutoff to inflation, we propose that
for cosmologically relevant scales, which have been amplified by inflation but
which are in the far IR at the end of inflation, no adiabatic 
subtraction should be performed.

One might argue, however, that this expansion scheme still produces a finite 
result and therefore 
provides a way to renormalize the IR modes. 
In fact, we find this argument not 
convincing, as there are different schemes 
to renormalize the IR modes (for example, see the recent paper 
\cite{Janssen}). In a way, we have also shown this by presenting 
different counterterms which agree in the far UV but not in the IR. This 
reflects the well known result that, in the far UV, where space-time curvature 
becomes negligible, the physical spectrum is independent of the regularization 
scheme, and that the UV singularity structure of the two-point function is 
always of the Hadamard form. 
Furthermore, since inflation has not started 
in the infinite past, there is a natural infrared cutoff, namely the horizon 
scale at the beginning of inflation.
We therefore conclude that one should 
not subtract the adiabatic contribution in the IR in realistic 
inflationary models.

Even though the adiabatic calculation does not apply in the IR, 
it is interesting to note that, the adiabatic counterterms become much 
smaller than the unrenormalized spectrum in the IR, when 
computed without any expansion, as in Eqs. (\ref{PZfull}) 
and (\ref{PTfull}). As mentioned above, these expressions show a singularity 
for $v^2\simeq 2$. This is not relevant, from our point of view, as we claim 
that the subtraction should be performed at the end of inflation.
At the end of inflation, however, all cosmologically relevant 
scales are in the far IR, hence the adiabatic subtraction, which 
is a possible prescription for the far UV does not 
affect the associated spectra. This is the main conclusion of this work.

The adiabatic subtraction does, however provide a clean means to derive 
the shape of the physical spectrum in the UV, where it actually tends to zero:
at any given time, fluctuations with $v>1$ are significantly suppressed
by the adiabatic counterterm. In this sense the adiabatic subtraction
provides a UV cutoff of the spectrum which is roughly given by the
scale $k_{UV}$ which reaches $v=1$ at the end of inflation, $k_{UV} =a_fH_f$.

The reason why it is usually sensible to compute $P_\zeta$ and $P_t$ at the 
Hubble exit, $v\simeq 1$ instead of evaluating them at the end of inflation, 
is that we have simple and sound formulae  for them, which are valid 
``inside the Hubble scale'', while the growing modes of the 
perturbations are nearly constant ``outside the Hubble scale''.
Therefore, in general we do not  need to calculate their evolution until the 
end of inflation. This is different for the adiabatic  counterterm $P^{(2)}$: as we have shown, this term 
becomes strongly time-dependent, and  decreases with $v$ in the IR, if 
extended in this regime without any further expansion. 
Therefore, it seems reasonable to perform the adiabatic subtraction
at the end of inflation, or at least far in the IR, where, however, it becomes 
irrelevant for all scales of cosmological interest.

\begin{center}{\large\bf
Acknowledgment }
\end{center}

We thank  Ivan Agull\'o, Jos\'e Navarro-Salas, Gonzalo J. Olmo, and Leonard 
Parker for interesting and clarifying discussions. GM also thanks
Alexei A. Starobinsky for useful discussions. This work is supported by the 
Swiss National Science Foundation.

\appendix


\section{Definitions}
\label{defin}

\noindent In Section \ref{three}, we present the renormalized power spectrum 
of curvature
and tensor perturbations, expanded with respect to the slow-roll parameters. 
The three
functions $f$, $g$, and $f_t$  appearing in these expressions are defined by 

\begin{figure}[!h]
\includegraphics[width=0.46\textwidth]{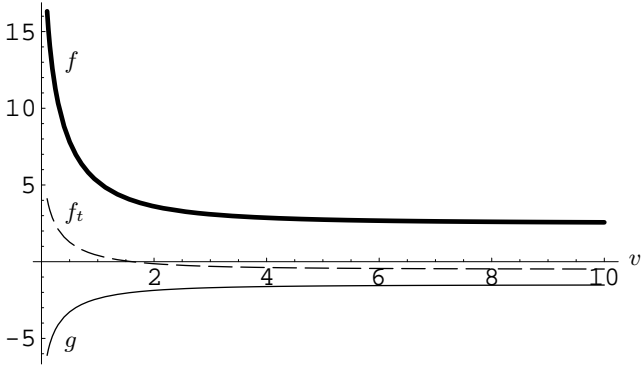}
\rput[tl](0,1.6){$v$}
\rput[tl](-7.5,4.3){$f$}
\rput[tl](-7.5,2.3){$f_t$}
\rput[tl](-7.5,0.5){$g$}
\caption{Plots of the behaviour of $f$ (thick line), $g$ (solid line) and
$f_t$ (dashed line)
in function of $v=k/(a H)$ in the range from $1/10$ to $10$.}
\label{figFunction}
\end{figure}

\begin{widetext}
\ba
f(v) &=& -2-3\sqrt{2 \pi}\, v^{5/2}\left\{\cos(v) \left[
\frac{\partial}{\partial \nu} J_\nu(v)\right]_{\nu=3/2}+\sin(v) 
\left[\frac{\partial}{\partial \nu} J_\nu(v)\right]_{\nu=3/2}\right\}+
\nonumber \\
 & & +3\sqrt{2 \pi}\, v^{3/2} \left\{\sin(v) \left[
\frac{\partial}{\partial \nu} J_\nu(v)\right]_{\nu=3/2}-\cos(v) 
\left[\frac{\partial}{\partial \nu} J_\nu(v)\right]_{\nu=3/2}\right\}\ ,
\label{fv} 
\ea
\ba
g(v) &=& \sqrt{2 \pi}\, v^{5/2}\left\{\cos(v) \left[
\frac{\partial}{\partial \nu} J_\nu(v)\right]_{\nu=3/2}+\sin(v) 
\left[\frac{\partial}{\partial \nu} J_\nu(v)\right]_{\nu=3/2}\right\}+
\nonumber \\
& & -\sqrt{2 \pi}\, v^{3/2} \left\{\sin(v) \left[
\frac{\partial}{\partial \nu} J_\nu(v)\right]_{\nu=3/2}-\cos(v) 
\left[\frac{\partial}{\partial \nu} J_\nu(v)\right]_{\nu=3/2}\right\}\ ,
\label {gv} 
\ea
\ba
f_t(v) &=& -2-\sqrt{2 \pi}\, v^{5/2}\left\{\cos(v) \left[
\frac{\partial}{\partial \nu} J_\nu(v)\right]_{\nu=3/2}+\sin(v) 
\left[\frac{\partial}{\partial \nu} J_\nu(v)\right]_{\nu=3/2}\right\}+
\nonumber \\
& & +\sqrt{2 \pi}\, v^{3/2} \left\{\sin(v) \left[
\frac{\partial}{\partial \nu} J_\nu(v)\right]_{\nu=3/2}-\cos(v) 
\left[\frac{\partial}{\partial \nu} J_\nu(v)\right]_{\nu=3/2}\right\}\,.
\label {ftv}
\ea
\end{widetext}
These functions are plotted in the range $1/10<v<10$ 
in Fig. \ref{figFunction}.

\section{Spectral indices}
\label{spectra}

\noindent We consider the case $V=m^2 \phi^2/2$. During slow-roll, we can 
neglect the 
terms $\dot\phi^2$ in Eq.\  (\ref{Fried}) and $\ddot\phi$ in Eq.\  
(\ref{eqmotion1}). It follows that $H\simeq H_i+\dot H t$, 
where $H_i$ is the initial value of the Hubble factor, and 
$\dot H\simeq-m^2/3$. Thus, the scale factor satisfies the equalities
\be
\ln {a(t)\over a_i}=\left(H_it-{m^2\over 6}t^2\right)= 
{3\over 2m^2}\left[H_i^2-H^2(t)\right]\ ,
\label{afactor}
\ee
where $a_i$ is its initial value. If we assume that inflation finishes 
approximately when $H(t)\simeq 0$, it follows that
\be
N\equiv\ln \frac{a_f}{a_N}=\frac{3}{2} \frac{H_N^2}{m^2}\ , \quad a_N=e^{-N} 
a_f\ ,
\ee
where $H_N$ and $a_N$ are the values of the Hubble and scale factors 
$N$ e-folds before the end of inflation.
Thus, we can write the momentum $k_N$, 
associated to the mode that exit the Hubble scale at $N$ 
$e$-folds before the end of the inflation, as 
\be
k_N\equiv a_N H_N=
m \left(\frac{2}{3}N\right)^{1/2} a_i 
\exp \left(\frac{3H_i^2}{2m^2}-N\right)\ .
\label{KN}
\ee
 Let $s$ be $v$ or $c$, according to whether we are dealing with the UV or IR 
 respectively. 
Let $t_s$ the time when 
\be
k_N=s a(t_s)H(t_s)\,,
\label{EqkN}
\ee
with the help of Eqs.\ (\ref{afactor}) and (\ref{KN}), we find a quadratic 
equation in $t_s$, which gives
\be
t_s={3H_i\over m^2}-{\sqrt{6}\over m}\sqrt{N+\ln s}\ ,
\ee
where we have replaced the slowly varying function $H(t_s)$ with the constant 
value $H_N$ in order to obtain an  analytical solution of Eq.~(\ref{EqkN}).
Then, as $H=H_i-m^2t/3$, we find
\be
H(t_s)=m\sqrt{{2\over 3}\left(N+\ln s \right)}\ ,
\ee
from which follows that
\be
\epsilon_s=\eta_s=2\left(\frac{M_{\rm pl}}{\phi_s}\right)^2 = 
\frac{m^2}{3H_s^2}
= \frac{1}{2 (N+ \ln s)}\ .
\ee

A possible way to evaluate $n_s$ is to express, at  leading order,  
the derivative with respect to $\ln k$ as
\be
\frac{d}{d \ln k}\simeq-\frac{d}{d N}\ ,
\ee
which yields, in both regimes, 
\be
n_s^{(UV)}=1-4 \epsilon_v\ , \quad n_s^{(IR)}=1-4 \epsilon_c\ .
\ee
In the UV, we obtain the standard result, 
while in the IR we have a slightly different expression.
In fact, even if the relation between $r$ and $n_s$ is the same in the two
regimes, namely $r=4-4 n_s$, in the IR we have
$n_s=1-2/(N+\ln c)$, while in the UV  we have $n_s=1-2/(N+\ln v)$. 
This difference is, however, quite small since $N\gg 1$ for scales which exit
during slow-roll.

\end{document}